\documentclass[journal]{IEEEtran}
\usepackage[normalem]{ulem}
\usepackage{color}

\ifCLASSINFOpdf
  \usepackage[pdftex]{graphicx}
\else
\fi
\usepackage{amsmath}
\usepackage{amsfonts}
\usepackage{algorithm}
\usepackage{algorithmic}
\begin{document}
%
\title{Synthetic Aperture Radar Image Formation \\ with Uncertainty Quantification}
%
%
%

\author{Victor~Churchill
        ~and~Anne~Gelb
\thanks{V. Churchill and A. Gelb are with the Department
of Mathematics, Dartmouth College, Hanover,
NH, 03755 USA e-mail: Victor.A.Churchill.GR@dartmouth.edu.  This work is partially supported by grants  NSF-DMS \#1502640, NSF-DMS \#1912685, and AFOSR \#FA9550-18-1-0316.}
}

\maketitle

\begin{abstract}
Synthetic aperture radar (SAR) is a day or night any-weather imaging modality that is an important tool in remote sensing. Most existing SAR image formation methods result in a maximum a posteriori image which approximates the reflectivity of an unknown ground scene. This single image provides no quantification of the certainty with which the features in the estimate should be trusted. In addition, finding the mode is generally not the best way to interrogate a posterior. This paper addresses these issues by introducing a sampling framework to SAR image formation. A hierarchical Bayesian model is constructed using conjugate priors that directly incorporate coherent imaging and the problematic speckle phenomenon which is known to degrade image quality. Samples of the resulting posterior as well as parameters governing speckle and noise are obtained using a Gibbs sampler. 
These samples may then be 
used to compute estimates, and also to derive other statistics like variance which aid in uncertainty quantification. The latter information is particularly important in SAR, where ground truth images even for synthetically-created examples are typically unknown. An example result using real-world data shows that the sampling-based approach introduced here to SAR image formation provides parameter-free estimates with improved contrast and significantly reduced speckle, as well as unprecedented uncertainty quantification information.
\end{abstract}


%
\IEEEpeerreviewmaketitle

\section{Introduction}\label{chap:introduction}

\IEEEPARstart{S}{ynthetic} aperture radar (SAR) is a widely-used imaging technology for surveillance and mapping. Because SAR is capable of all-weather day-or-night imaging, it overcomes several challenges faced by optical imaging technologies and is an important tool in modern remote sensing, \cite{jakowatz2012spotlight}. Applications for SAR include areal mapping and analysis of ground scenes in environmental monitoring, remote mapping, and military surveillance, \cite{andersson2012fast}. To collect SAR data, an airborne sensor traverses a flight path, periodically transmitting an interrogating waveform toward an illuminated region of interest. The emitted energy pulses impinge on targets in the illuminated region that scatter electromagnetic energy back to the sensor. The sensor measures and processes the reflected signal. The demodulated data, called a phase history, is passed on to an image formation processor. This paper concerns image formation, which produces a reconstruction of the two-dimensional electromagnetic reflectivity function of the illuminated ground scene from SAR phase history data. For an overview of SAR and basic image formation techniques including back projection, see e.g. \cite{jakowatz2012spotlight,gorham2010sar}. 

Several issues pose challenges to forming artifact- and noise-free SAR images. 
First, large images and data prohibit use of traditional matrix-based methods for linear inverse problems, as even storing dense matrices of this size is problematic. In addition, SAR is a coherent imaging system, 
and is therefore affected by speckle, a multiplicative-noise-like phenomenon which causes grainy-looking images, \cite{argenti2013tutorial}. The result of most existing methods is a single image, typically a maximum \emph{a posteriori} (MAP) point estimate, that approximates the unknown ground truth. There are several issues with MAP estimates. First, the maximum is not categorically representative of the posterior -- in general sampling is a better way to interrogate a density. Second, MAP estimates rely on user-defined parameters, but estimating these quantities would impose less bias. 
Finally, the MAP prediction is not probabilistic, and therefore provides no estimate of the statistical confidence with which we can trust the features in the image. 
In a Bayesian formulation, an entire joint density estimates the unknown quantities. We show that this more complete information is particularly important in SAR, where even in synthetically-created examples the ground truth reflectivity is unknown.

In this paper we therefore develop a parameter-free sampling-based SAR image formation framework with uncertainty quantification. 
A hierarchical Bayesian network structure, \cite{bardsley2012mcmc}, allows our model to directly incorporate coherent imaging and the problematic speckle phenomenon into the prior. A joint posterior density estimate is computed for the image, the noise parameter, and the speckle parameter. 
By encouraging sparsity in the magnitude of the image we are able to reduce speckle and increase contrast. It is important to note that all parameters in the model are prescribed, requiring no user input. Conjugate priors are used so that the resulting posterior can be efficiently sampled using a Gibbs sampler. 
The samples obtained from the posterior density can be used in a variety of ways, including to compute estimates of the unknown reflectivity image as well as the noise parameter and crucially the speckle parameter, which typically requires post-processing to quantify, \cite{argenti2013tutorial}. In addition, samples provide variance and other statistics to aid in uncertainty quantification.

The rest of this paper is organized as follows. Section \ref{sec:model} describes the hierarchical Bayesian model which directly incorporate coherent imaging, speckle, and sparsity using conjugate priors. Section \ref{sec:sampling} outlines an efficient sampling method for the resulting posterior. Section \ref{sec:results} shows an example. 
The results suggest that the proposed method provides estimates with significantly better contrast and drastically reduced speckle compared with other methods, in addition to the unprecedented UQ. Finally, Section \ref{sec:conclusion} gives a summary of the paper and directions for future research.

\section{Hierarchical Bayesian Model for SAR}\label{sec:model}
We now formulate  a hierarchical Bayesian model and joint posterior.

\subsection{Discrete model}


The measured SAR phase history data can be modeled as a continuous non-uniform Fourier transform of the reflectivity function, \cite{sanders2017composite}. Here we use the discretized model
\begin{align}\label{eq:discretemodel}
\mathbf{\hat{f}} = \mathbf{F}\mathbf{f}+\mathbf{n},
\end{align}
where $\mathbf{\hat{f}}\in\mathbb{C}^{M}$ is the vertically-concatenated phase history data, $\mathbf{F}\in\mathbb{C}^{M\times N}$ is a two-dimensional discrete non-uniform Fourier transform matrix, and the vector $\mathbf{f}\in\mathbb{C}^{N}$ is the vertically-concatenated unknown reflectivity image matrix. Finally, $\mathbf{n}\in\mathbb{C}^{M}$ represents model and measurement error. Note that by using the discrete Fourier transform in (\ref{eq:discretemodel}) we introduce both aliasing error and the Gibbs phenomenon. 

The objective is to infer $\mathbf{f}$ from $\mathbf{\hat{f}}$. Observe that while SAR images are complex-valued, usually only the magnitude is viewed while the phase is omitted.  However, recovering the phase is also important, \cite{moore2018characterization}, and should not be neglected. 
It is also common to modify \eqref{eq:discretemodel} to include autofocusing for the purpose of phase error reduction, \cite{scarnati2018joint}. While such modifications are not a primary concern in this investigation, they can be incorporated into the proposed method in a straightforward manner.


Throughout this paper we assume $\mathbf{n}$ is complex circularly-symmetric white Gaussian noise. That is, $\mathbf{n}_i\sim\mathcal{CN}(0,\beta^{-1})$ i.i.d. for all $i$, where 
$\beta^{-1}>0$ is the noise variance. 
This yields the Gaussian likelihood function
\begin{align}\label{eq:likelihood}
p(\mathbf{\hat{f}}| \mathbf{f}, \beta) \propto \beta^{M} \exp(-\beta||\mathbf{\hat{f}}-\mathbf{F}\mathbf{f}||^2),
\end{align}
which measures the goodness of fit of the model \eqref{eq:discretemodel}, where $||\mathbf{g}||^2 = \mathbf{g}^H\mathbf{g}$ with $\mathbf{g}^H$ the conjugate transpose of $\mathbf{g}$.

\subsection{Hierarchical prior}

With the likelihood given by \eqref{eq:likelihood}, the next step in computing the posterior is to specify a prior for the latent variable $\mathbf{f}$. 
We use the fact that SAR images are affected by the speckle phenomenon as a prior. Speckle, which occurs in all coherent imaging and is  often  misidentified and mischaracterized as noise, causes a complicated granular pattern of bright and dark spots throughout an image, \cite{jakowatz2012spotlight}. Although speckle is in fact signal and {\em not} noise, it nonetheless degrades the image quality by lowering the contrast, and hence it is desirable to remove it. While speckle reduction is often tackled using denoising techniques, here instead we directly incorporate the speckle into the image formation model, so that it is properly characterized as part of the signal. 
Specifically we employ the fully-developed speckle model which is based on the assumption that the spatial resolution dimension is considerably larger than the wavelength and the illuminated surface is rough enough, \cite{dong2014sar}.  Details of this popular model are described below.

Assume $\text{Re}(\mathbf{f}_i)$ and $\text{Im}(\mathbf{f}_i)$ are respectively i.i.d. Gaussian with variance $\boldsymbol{\alpha}_{i}^{-1}$. That is,
$\text{Re}(\mathbf{f}_i),\text{Im}(\mathbf{f}_i)\sim\mathcal{N}(0,\boldsymbol{\alpha}^{-1}_i)$.
By independence, 
$\text{Re}(\mathbf{f}),\text{Im}(\mathbf{f})\sim\mathcal{N}(\mathbf{0},\text{diag}(\boldsymbol{\alpha}^{-1}))$.
This is conveniently encoded by $\mathbf{f}\sim\mathcal{CN}(\mathbf{0},\text{diag}(\boldsymbol{\alpha}^{-1}))$  which means that  $\mathbf{f}$ is circularly-symmetric complex Gaussian with density
\begin{align}\label{eq:jointprior}
p(\mathbf{f}|\boldsymbol{\alpha}) \propto \prod_{i=1}^N \boldsymbol{\alpha}_i \exp(-||\sqrt{\boldsymbol{\alpha}}\odot\mathbf{f}||^2),
\end{align}
where $\odot$ is elementwise multiplication. Thus we see that the prior on the magnitude $|\mathbf{f}_i|=\sqrt{\text{Re}(\mathbf{f}_i)^2+\text{Im}(\mathbf{f}_i)^2}$ is a Rayleigh probability distribution with mean proportional to $\boldsymbol{\alpha}_i^{-1}$. This is the standard specification for fully-developed speckle. Because the changes in the magnitude of each pixel $|\mathbf{f}_i|$ is proportional to $\boldsymbol{\alpha}_{i}^{-1}$, the speckle phenomenon has also been modeled as a multiplicative noise, \cite{dong2014sar}. While efforts to reduce speckle abound, \cite{argenti2013tutorial}, here we address the speckle directly by including it in our model with the prior \eqref{eq:jointprior}, and later estimating the associated speckle parameter $\boldsymbol{\alpha}^{-1}$ through sampling rather than post-image-formation techniques. Note that by parametrizing $\mathbf{f}$ with $\boldsymbol{\alpha}$ we are actually doubling the number of parameters in this model, which clearly provides a \emph{computational} challenge  (but not a methodological one). We will return to this issue later. 

As mentioned previously, using the MAP estimate as the solution is limiting -- first because it may not be representative of the posterior and second  
because it provides no information about the statistical confidence of the estimate of each recovered pixel value, or in any other recovered features of the image, \cite{nagy2002image}.  
Finally, the regularization parameters for both the cost function and prior in the MAP estimate approach (analogous to $\beta^{-1}$ and $\boldsymbol{\alpha}^{-1}$ here) are {\em user-specified}.  However they are truly unknown and therefore should be inferred. 
For these reasons we seek the joint posterior $p(\mathbf{f},\boldsymbol{\alpha},\beta | \mathbf{\hat{f}})$. More importantly, we will be estimating an entire density for the speckle parameter $\boldsymbol{\alpha}^{-1}$, which will lend clarity when determining whether or not the speckle reduction techniques are actually working.\footnote{Without a reference ground truth image, speckle statistics are typcically only estimated from small regions of already formed images, \cite{argenti2013tutorial}.} In order to calculate this posterior, we must define priors on $\boldsymbol{\alpha}$ and $\beta$. 
To this end, we first invoke a conjugate Gamma prior for $\beta$. That is, $\beta\sim\Gamma(c,d)$ with probability density function
\begin{align}\label{eq:betaprior}
p(\beta|c,d) \propto \beta^{c-1}\exp(-d\beta).
\end{align}
Similarly a conjugate Gamma prior is invoked on each element of $\boldsymbol{\alpha}$, i.e. $\boldsymbol{\alpha}_{i}\sim\Gamma(a,b)$. 
By independence, $\boldsymbol{\alpha}\sim\Gamma(a,b)$ with
\begin{align}\label{eq:priorA}
p(\boldsymbol{\alpha}|a,b) \propto & \prod_{i=1}^N \boldsymbol{\alpha}^{a-1}_i \exp\left(-b\sum_{i=1}^N\boldsymbol{\alpha}_i\right).
\end{align}
Note the dependence of \eqref{eq:betaprior} and \eqref{eq:priorA} on parameters $a$, $b$, $c$, and $d$, which as in \cite{bardsley2012mcmc,tipping2001sparse} are chosen rather than inferred. In \cite{bardsley2012mcmc}, analogous parameters in a real-valued model are chosen to reflect the uncertainty in the latent variable, making the prior uninformative. On the other hand in \cite{tipping2001sparse} $a=b=c=d=0$, resulting in an improper prior $p(\mathbf{f}_i)\sim1/|\mathbf{f}_i|$, which is peaked at zero and hence encourages sparsity.\footnote{To ensure numerical robustness in our implementation, we choose these parameters to be machine precision rather than $0$ in Algorithm \ref{alg:sample}.} Sparsity-promoting image formation methods for SAR is popular, see e.g.~\cite{scarnati2018joint,dong2014sar,sanders2017composite,archibald2016image,ccetin2001feature}. 

For purpose of comparison, then, in the figures that follow we also show images formed using two popular sparsity-encouraging image formation methods: (1) $\ell_1$ regularization, where sparsity is promoted in the image itself; and (2) total variation (TV) regularization, where sparsity is promoted in the gradients in  the image. Both methods have been extensively applied in SAR (see e.g. \cite{scarnati2018joint,dong2014sar,sanders2017composite,archibald2016image,ccetin2001feature}).
Importantly, choosing $a, b, c$ and $d$ in this way removes any need for user-defined parameters in this model.

\subsection{Posterior computation}

The form of the posterior is achieved through the hierarchical Bayesian model described above, where the model parameters $\mathbf{f}$ and $\beta$ are given priors with prior parameters $\boldsymbol{\alpha}$, $c$, and $d$), referred to as hyperparameters. Moving up to the final level of hierarchy in this model, the hyperparameter $\boldsymbol{\alpha}$ is given a prior (called a hyperprior) with hyperhyperparameters $a$ and $b$.
By Bayes' theorem, the joint posterior for $\mathbf{f}$, $\boldsymbol{\alpha}$, and $\beta$ is
\begin{align}\label{eq:posterior}
p(\mathbf{f},\boldsymbol{\alpha},\beta|\mathbf{\hat{f}}) &\propto p(\mathbf{\hat{f}}|\mathbf{f},\beta)p(\beta|0,0)p(\mathbf{f}|\boldsymbol{\alpha})p(\boldsymbol{\alpha}|0,0) \\
&\propto \beta^{M-1}\exp(-\beta||\mathbf{\hat{f}}-\mathbf{F}\mathbf{f}||^2-||\sqrt{\boldsymbol{\alpha}}\odot\mathbf{f}||^2). \nonumber
\end{align}
The image formation algorithm in Section \ref{sec:sampling} will require the individual posteriors for each latent variable.
Because of the conjugate priors used, these can be found analytically. We have
\begin{subequations}
\begin{align}
\label{eq:bayesf}
p(\mathbf{f}|\mathbf{\hat{f}},\boldsymbol{\alpha},\beta) &\propto \exp(-\beta||\mathbf{\hat{f}}-\mathbf{F}\mathbf{f}||^2-||\sqrt{\boldsymbol{\alpha}}\odot\mathbf{f}||^2)
\end{align}
\begin{align}
\label{eq:bayesa}
p(\boldsymbol{\alpha}|\mathbf{\hat{f}},\mathbf{f},\beta) 
&\propto \exp(-||\sqrt{\boldsymbol{\alpha}}\odot\mathbf{f}||^2 )\\
\end{align}
\begin{align}
\label{eq:bayesb}
p(\beta|\mathbf{\hat{f}},\mathbf{f},\boldsymbol{\alpha}) &\propto \beta^{M-1} \exp(-\beta||\mathbf{\hat{f}}-\mathbf{F}\mathbf{f}||^2).
\end{align}
\label{eq:bayesequations}
\end{subequations}
Therefore each latent variable can be sampled from
\begin{subequations}
\begin{align}
\mathbf{f}|\mathbf{\hat{f}},\boldsymbol{\alpha},\beta &\sim \mathcal{CN}(\beta\Sigma\mathbf{F}^H\mathbf{\hat{f}},\Sigma)\label{eq:fpost}
\end{align}
\begin{align}
\boldsymbol{\alpha}|\mathbf{\hat{f}},\mathbf{f},\beta&\sim \Gamma(1,\mathbf{f}\odot\bar{\mathbf{f}})\label{eq:alphapost}
\end{align}
\begin{align}
\beta|\mathbf{\hat{f}},\mathbf{f},\boldsymbol{\alpha}&\sim \Gamma(M,||\mathbf{\hat{f}}-\mathbf{F}\mathbf{f}||^2)\label{eq:betapost}.
\end{align}
\label{eq:fabpost}
\end{subequations}
with $\Sigma=(\beta\mathbf{F}^H\mathbf{F}+\text{diag}(\boldsymbol{\alpha}))^{-1}$.

\normalsize
\section{Sampling-based SAR Image Formation}\label{sec:sampling}

Following the approach in \cite{bardsley2012mcmc}, we now set up a sampling-based image formation procedure to obtain approximate samples from \eqref{eq:posterior}. 
Since the density is not described by a known family of probability distributions, 
it cannot be directly sampled. 
However, because of the conjugate prior structure, we can easily apply a Gibbs sampler, \cite{geman1984stochastic}, which obtains approximate samples from the joint posterior \eqref{eq:posterior} by sequentially sampling the individual posteriors for each variable given in \eqref{eq:fabpost}.  
As with other Markov chain Monte Carlo (MCMC) methods, Gibbs sampling creates a Markov chain of samples, each of which is correlated with the other samples.



In terms of efficiency, an issue occurs in sampling the individual posterior for $\mathbf{f}$ given by \eqref{eq:fpost}, where we must solve the following linear system determined by \eqref{eq:bayesf} for $\mathbf{f}$: 
\begin{align}
\beta\mathbf{F}^H\mathbf{F}\mathbf{f}+\boldsymbol{\alpha}\odot\mathbf{f} &= \beta\mathbf{F}^H(\mathbf{\hat{f}}+\mathbf{v}_1)+\mathbf{v}_2,\label{eq:linsys1}
\end{align}
with $\mathbf{v}_1\sim\mathcal{CN}(\mathbf{0},\mathbf{I}/\sqrt{\beta})$ and $\mathbf{v}_2\sim\mathcal{CN}(\mathbf{0},\text{diag}(\sqrt{\boldsymbol{\alpha}}))$, \cite{bardsley2012mcmc}. Even storing the dense matrices $\mathbf{F}$ and $\mathbf{F}^H$ in real-world problems is not practical. 
However, because $\mathbf{F}$ is a non-uniform discrete Fourier transform matrix, we can utilize existing libraries to quickly apply a non-uniform fast Fourier transforms (NUFFT), \cite{fessler2003nonuniform}. 
The NUFFT is performed by interpolating non-uniform Fourier mode quantities to a uniform grid so that the uniform FFT can be used, \cite{fessler2003nonuniform}. This is not without error of course, which mainly comes from the error accumulated when
``gridding'' non-uniform to uniform Fourier modes. Further work will be needed to meaningfully quantify this error for this application. For the current investigation, in order to apply $\mathbf{F}$ efficiently, we employ a unitary operation. This means that the left-hand-side operation of \eqref{eq:linsys1} can be approximately diagonalized as
\begin{align}
\beta\mathbf{F}^H\mathbf{F}\mathbf{f}+\boldsymbol{\alpha}\odot\mathbf{f} \approx (\beta+\boldsymbol{\alpha})\odot\mathbf{f},
\label{eq:linsys1approx}
\end{align}
 which can now be efficiently inverted using simple elementwise division. Clearly using the right hand side in \eqref{eq:linsys1approx} introduces additional error, along with that from modifying the non-uniform modes in order to make them conform with a uniform grid, oscillations due to the Gibbs phenomenon, and model and measurement error. A potentially more accurate method would be to use elementwise division by $\beta+\boldsymbol{\alpha}$ as a preconditioner in a conjugate gradient descent scheme, however this would be far less efficient. 
By combining \eqref{eq:bayesequations}, \eqref{eq:fabpost}, and \eqref{eq:linsys1approx} we arrive at Algorithm \ref{alg:sample}, which produces 
$K$ samples for $\mathbf{f}$, $\boldsymbol{\alpha}$, and $\beta$, each of which are approximately drawn from the joint posterior. Notice that each sample requires two NUFFT applications.
\begin{algorithm}[h]
\caption{Sampling from $p(\mathbf{f},\boldsymbol{\alpha},\beta|\mathbf{\hat{f}})$}
\label{alg:sample}
\begin{algorithmic}
\STATE{Initiate $\mathbf{f}^0$, $\boldsymbol{\alpha}^0$, $\beta^0$;}
\FOR{$k=1$ to $K$}
\STATE{Draw $\mathbf{v}_1\sim\mathcal{CN}(\mathbf{0},\mathbf{I}/\sqrt{\beta})$ and $\mathbf{v}_2\sim\mathcal{CN}(\mathbf{0},\text{diag}(\sqrt{\boldsymbol{\alpha}}))$;}

\STATE{Compute $\mathbf{f}^{k+1} =(\beta^k\mathbf{F}^H(\mathbf{\hat{f}} +\mathbf{v}_1)+\mathbf{v}_2)\oslash (\beta^k+\boldsymbol{\alpha}^k)$;}
\STATE{Compute $\boldsymbol{\alpha}^{k+1}\sim\Gamma(1,\mathbf{f}^{k+1}\odot\overline{\mathbf{f}^{k+1}})$;}
\STATE{Compute $\beta^{k+1}\sim\Gamma(M,||\mathbf{\hat{f}}-\mathbf{Ff}^{k+1}||^2)$;}
\ENDFOR
\end{algorithmic}
\end{algorithm}

\subsection{Chain convergence}

It is generally unknown how quickly the chain formed in Algorithm \ref{alg:sample} will converge. 
While there are several heuristic approaches available, here we follow \cite{gelman2013bayesian} to determine chain convergence. Multiple chains are computed using randomly chosen starting points based on the observation that the variance within a single chain will converge faster than the variance between chains. A statistic is computed for each element of each latent variable, the value of which is a measure of convergence of that individual parameter. The derivation of this statistic closely follows \cite{bardsley2012mcmc}. Compute $n_r$ chains (in our implementation this is done in parallel) each of length $2n_s$, keeping only the latter $n_s$ samples. Let $\psi_{ij}$ denote the $i$th sample from the $j$th chain for a single parameter, and define
\begin{align*}
B &= \frac{n_s}{n_r-1} \sum_{j=1}^{n_r} \left(\bar{\psi}_{\cdot j} - \bar{\psi}_{\cdot\cdot}\right)^2,
\end{align*}
where $\bar{\psi}_{\cdot j}$ is the mean of the samples in the chain $j$, $\bar{\psi}_{\cdot\cdot}$ is the mean of the samples in every chain, and
\begin{align*}
W = \frac{1}{n_r}\sum_{j=1}^{n_r}s_j^2,\quad\text{with}\quad s_j^2 = \frac{1}{n_s-1}\sum_{i=1}^{n_s} \left(\psi_{i j} - \bar{\psi}_{\cdot j}\right)^2.
\end{align*}
Hence $B$ is a measure of the variance between the chains while $W$ is a measure of the variance within each individual chain. The marginal posterior variance $\text{var}(\psi|\mathbf{\hat{f}})$ is then estimated by
\begin{align}
\widehat{\text{var}}^+(\psi|\mathbf{\hat{f}}) = \frac{n_s-1}{n_s}W+\frac{1}{n_s}B,
\end{align}
which is an unbiased estimate under stationarity, \cite{gelman2013bayesian}. From this variance estimate, we compute the desired statistic
\begin{align}
\hat{R} = \sqrt{\frac{\widehat{\text{var}}^+(\psi|\mathbf{\hat{f}})}{W}},
\end{align}
which tends to $1$ from above as $n_s\rightarrow\infty$. Once $\hat{R}$ dips below $1.1$ for all sampled parameters, the $n_s n_r$ samples can together be considered samples from the posterior \eqref{eq:posterior}, \cite{gelman2013bayesian}. Note that other values can also be chosen as a tolerance for $\hat{R}$, \cite{bardsley2012mcmc}, but using $1.1$ seems reasonable when accounting for additional numerical errors. 
From the resulting $n_sn_r$ samples of $\mathbf{f}$, $\boldsymbol{\alpha}$, and $\beta$, a variety of sample statistics can be computed which describe the joint posterior density and help to quantify the uncertainty in the data.

\begin{figure}[t]
\centering
\includegraphics[width=.49\textwidth]{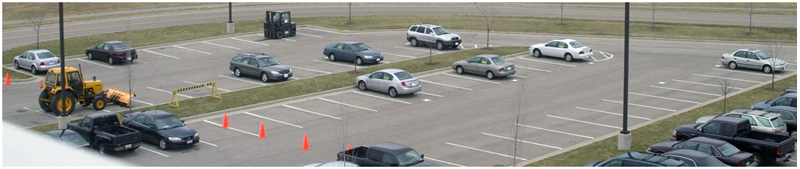}\\
\includegraphics[width=.24\textwidth]{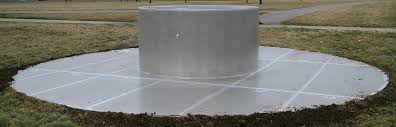}
\includegraphics[width=.24\textwidth]{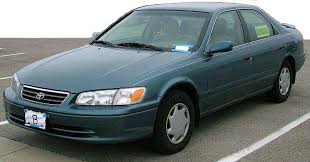}
\caption{Optical images of parking lot being imaged in GOTCHA dataset, \cite{casteel2007challenge}. Note scene contains a variety of calibration targets, such as primitive reflectors like the tophat shown, a Toyota Camry, forklift, and tractor.}
\label{fig:parking_lot}
\end{figure}

\section{Results}\label{sec:results}

We now provide a real-world example that demonstrates the robustness of Algorithm \ref{alg:sample} for SAR image formation. Since the ground truth reflectivity images is unknown, we are unable to compute standard error statistics such as the relative error.
This is the case even in synthetically-created SAR examples, where the true reflectivity is still unknown. Therefore, the unprecedented uncertainty quantification information the proposed method provides is all the more valuable, as it is able to quantify how much we should trust pixel values and structures in the image even in the absence of ground truth. Throughout, reflectivity images $\mathbf{f}$ are displayed in decibels (dB): $20\log_{10}\left(\frac{|\mathbf{f}|}{\max |\mathbf{f}|}\right)$, with a minimum of $-60$ dB and maximum of $0$ dB. Lesser or greater values are assigned the minimum or maximum.

\subsection{Data}

The GOTCHA Volumetric SAR Data Set consists of SAR phase history data of a parking lot scene collected at X-band with a 640 MHz bandwidth with full azimuth coverage at 8 different elevation angles with full polarization, \cite{casteel2007challenge}. This is a real-world SAR dataset captured by the Air Force Research Laboratory. The parking lot contains various vehicle targets including civilian vehicles, construction vehicles, calibration targets, primitive reflectors, and military vehicles. Figure~\ref{fig:parking_lot} shows optical images of the targets. 
The center frequency is 9.6GHz and bandwidth is 640MHz. This public release data has been used extensively for testing new SAR image formation methods. 

\begin{figure}[t]
\centering
\includegraphics[width=.24\textwidth]{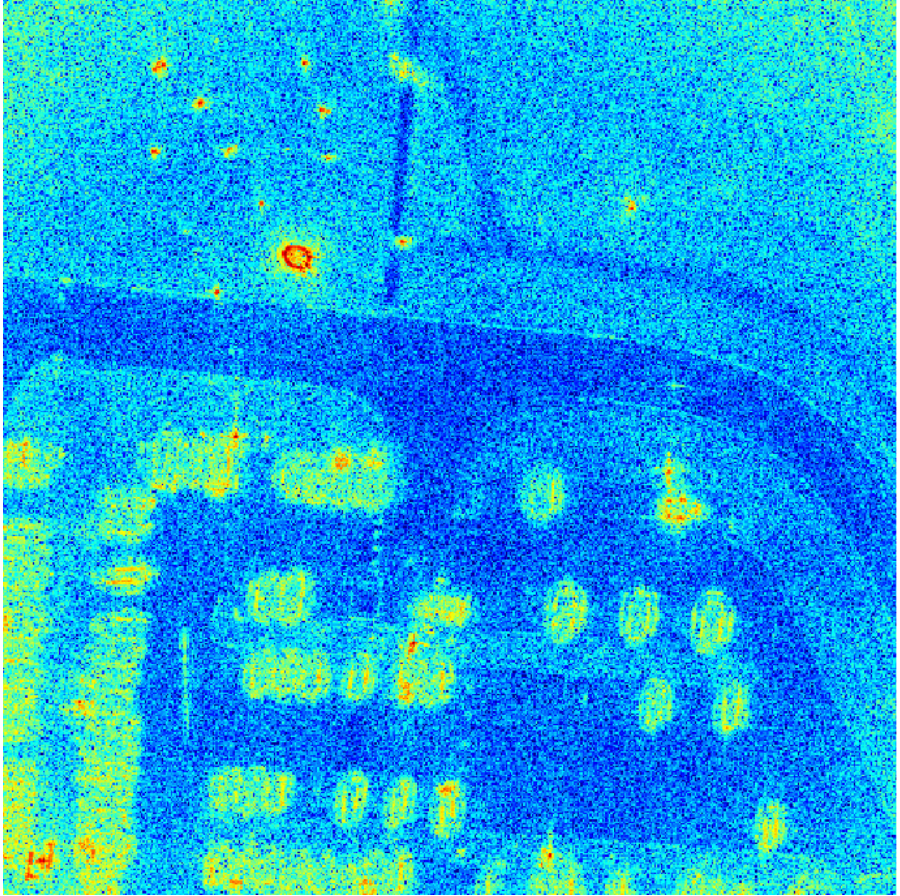}
\includegraphics[width=.24\textwidth]{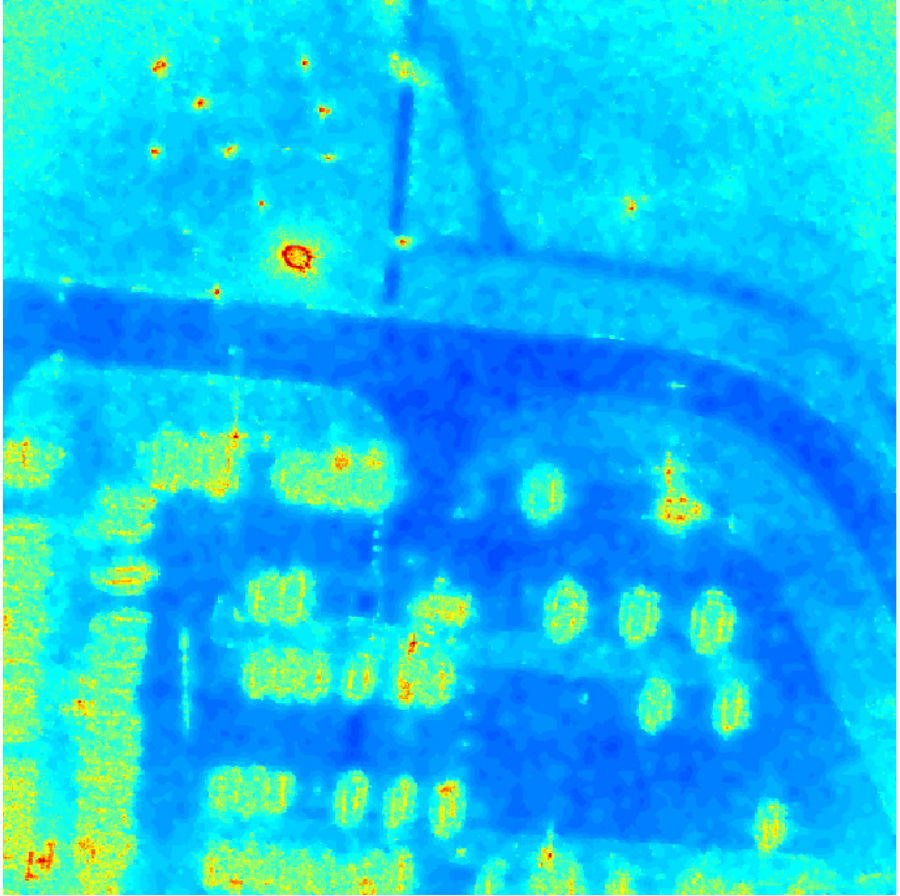}
\includegraphics[width=.24\textwidth]{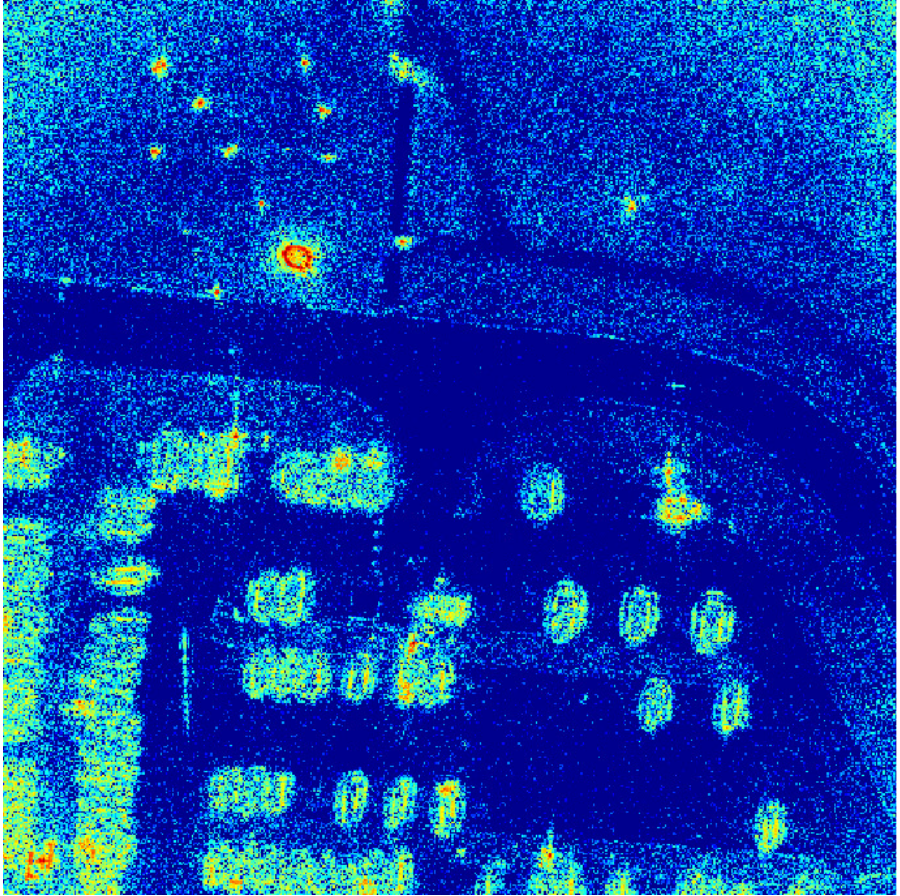}
\includegraphics[width=.24\textwidth]{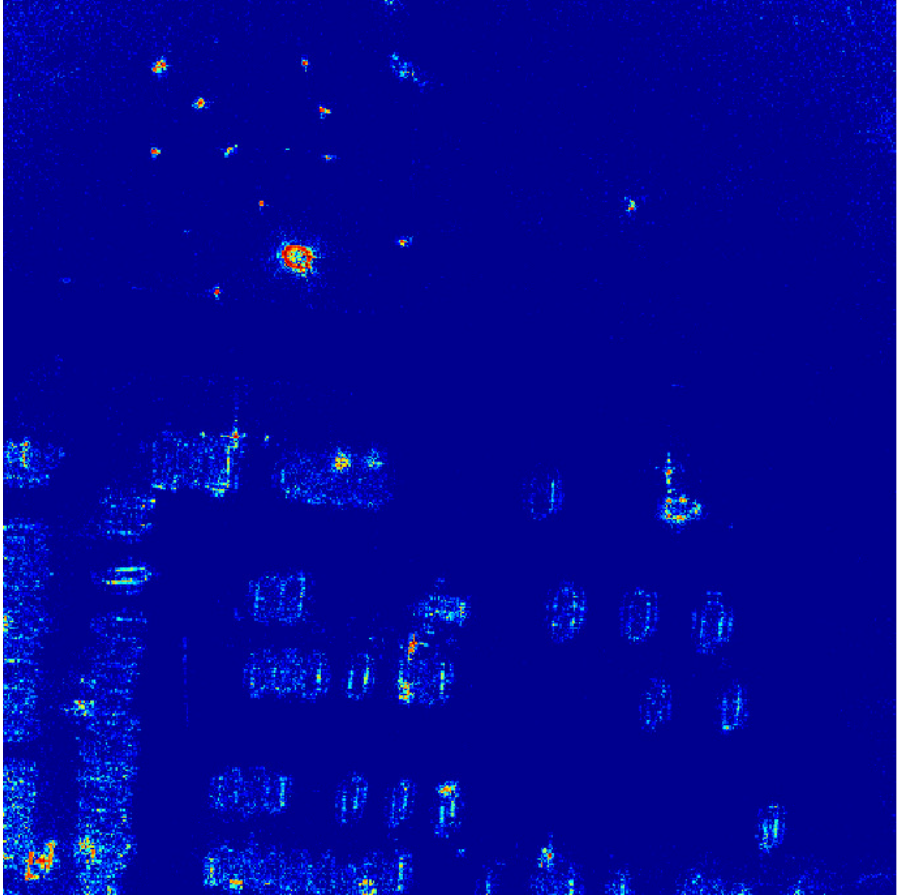}
\caption{Full images formed with (clockwise from top left) NUFFT; TV regularization; sparse sample mean; $\ell_1$ regularization.}
\label{fig:GOTCHA_compare}
\end{figure}

\subsection{Image estimate}\label{sec:examples}

Figure \ref{fig:GOTCHA_compare} compares images of the parking lot scene using an inverse NUFFT, $\ell_1$ regularization, total variation (TV) regularization, and the proposed sparsity-based sampling method. The inverse NUFFT image corresponds to a maximum likelihood estimate, minimizing a least squares cost function. Clearly this does little to reduce speckle and noise and results in a grainy image. The $\ell_1$ regularization scheme encourages sparsity in the estimate, yet it is evident that much of the speckle remains. The TV regularization perhaps does the best at removing speckle, however it leaves block-like artifacts in its place, making it difficult to distinguish between signal and background. Recall that TV regularization is essentially an image denoising model -- it aims to recover a piecewise constant image and also does not distinguish speckle from noise -- which may explain the results.  The images have $N=512^2$ pixels. Code from \cite{sanders-imaging} is used to wrangle the GOTCHA data and perform image formation for the comparison methods. 
The sampling-based method, which also uses sparsity-encouraging parameters, appears to retrieve a significantly better estimate than the other methods in terms of noise and speckle reduction, as well as contrast improvement.  There also appears to be no new artifacts, such as the block-like artifacts in the TV reconstruction.

The runtimes for each algorithm were $.03$ s for the NUFFT, $5.8$ s for the $\ell_1$ regularization, and $526$ s for the sampling method. Each image was formed on Polaris, a shared memory computer operated by Dartmouth Research Computing with 40 cores, 64-bit Intel processors, and 1 TB of memory. Using such a large machine was necessary in order to store the samples (here $n_r n_s=5\cdot1322$ for each of $3\times512^2+1$ parameters). 




\begin{figure}[t]
\centering
\includegraphics[width=.24\textwidth]{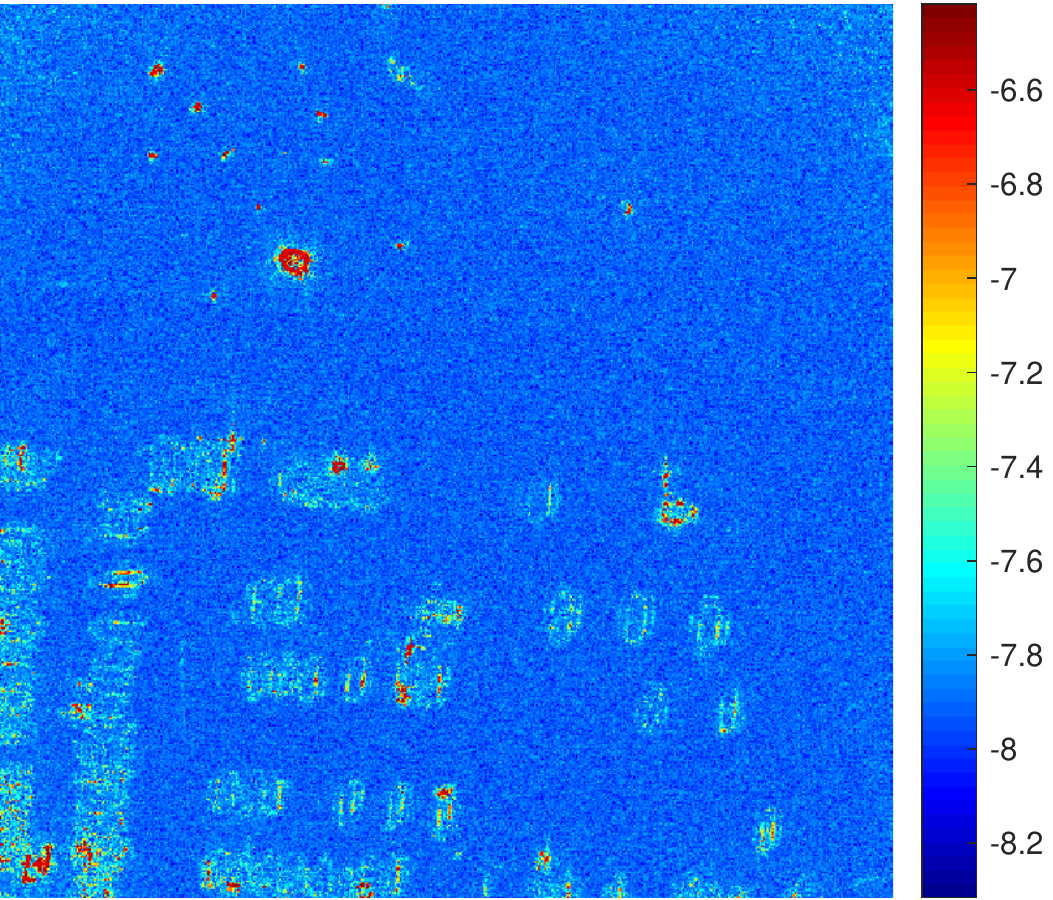}
\caption{$\log_{10}$ of the sample variance of $\mathbf{f}$.}
\label{fig:variance_compare}
\end{figure}

\subsection{Uncertainty quantification}

With the samples having been drawn, and image estimates computed, we now seek to visualize confidence information in order to provide important information about the certainty of these estimates. 
One way to do this is to look at the variance of the samples (or standard deviation) at each pixel. This can be helpful in forming a confidence estimate by acknowledging that roughly $2$ standard deviations from the mean contains $95\%$ of samples in a Gaussian distribution. Figure \ref{fig:variance_compare} shows $\log_{10}$ of the sample variance of $\mathbf{f}$. We see that high-magnitude pixels tend to vary more than low-magnitude pixels, which is indicative of tighter confidence intervals in the background, implying greater certainty that there are no targets in these regions. 
We can also display the $n_r n_s$ samples in a short movie. Using Twinkle, \cite{nagy2002image}, as inspiration, we are less confident in pixels or structures in the image that ``twinkle'' than those that are persistent throughout the movie. 



\begin{figure}[t]
\centering
\includegraphics[width=.24\textwidth]{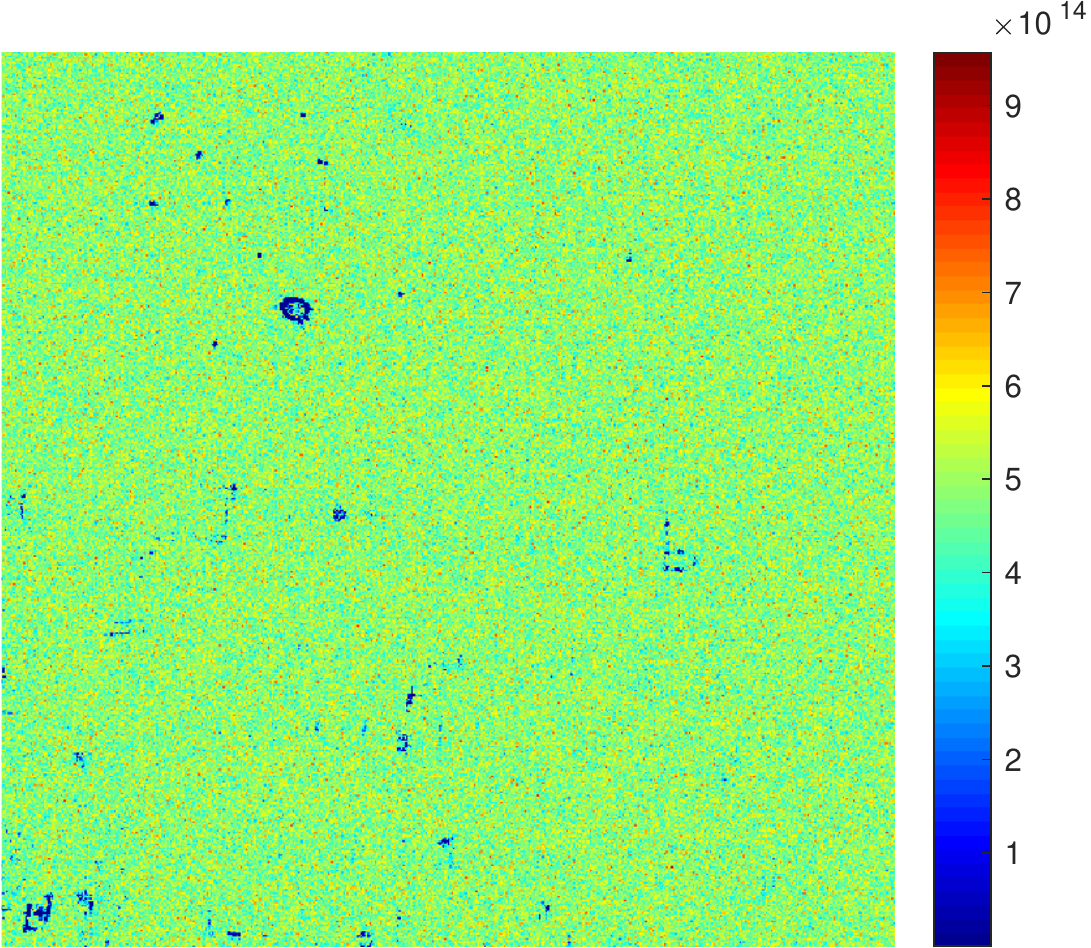}
\includegraphics[width=.24\textwidth]{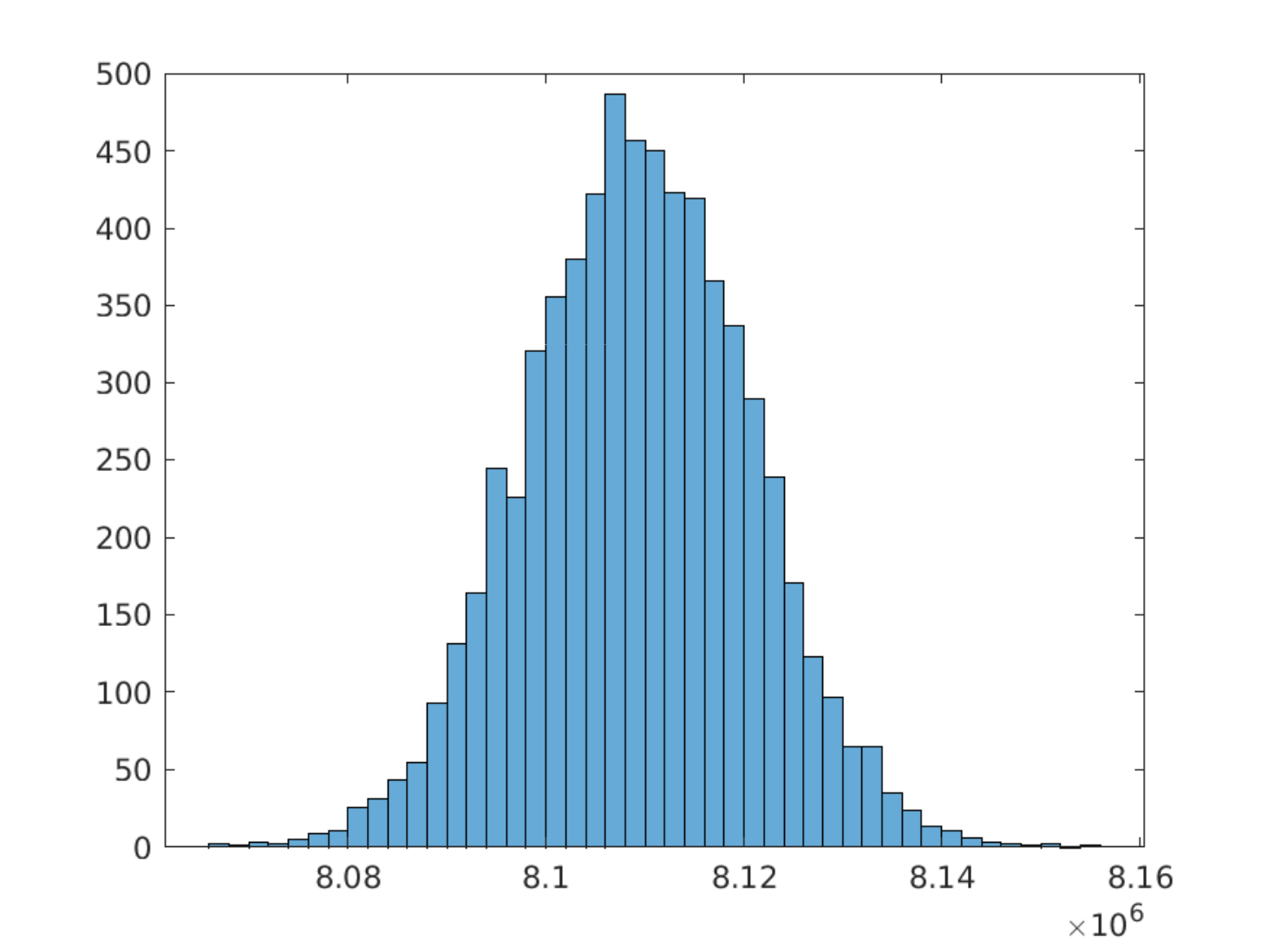}
\caption{(Left) Sample mean of $\boldsymbol{\alpha}$; (right) histogram of $\beta$ samples.}
\label{fig:parameter_compare}
\end{figure}

\subsection{Speckle and noise estimates}




Similar to the image estimates, the sampling-based image formation method produces samples of $\boldsymbol{\alpha}$, the parameter governing speckle, and $\beta$, the inverse noise variance. Figure \ref{fig:parameter_compare} shows the sample mean of $\boldsymbol{\alpha}$ as well as a histogram of the scalar $\beta$ samples. The sample mean for $\beta$ was $1.2331\times10^{-7}$. We also have that the reciprocal values of this image provide an estimate for the mean speckle parameter. There are several observations that can be made from the $\boldsymbol{\alpha}$ estimate. 
Many features of the reflectivity image are visible in this estimate. 
Recall that the magnitude of each pixel $|\mathbf{f}_i|$ is Rayleigh distributed with mean proportional to $\boldsymbol{\alpha}_{i}^{-1}$, hence changes in the magnitude of each pixel $|\mathbf{f}_i|$ are proportional to $\boldsymbol{\alpha}_{i}^{-1}$. 
There is practically no speckle (on the order $10^{-14}$) except at the various high-magnitude target reflectivities where the fully-developed speckle model is no longer appropriate, matching the speckle reduction we saw in the image estimate in Figure \ref{fig:GOTCHA_compare}. This confirms that the sparsity-encouraging measures taken effectively reduced speckle. 
In addition to these estimates, we can also perform similar uncertainty quantification analysis for the  $\boldsymbol{\alpha}$ by looking at the variance image or sample movie as well. 

\section{Conclusion}\label{sec:conclusion}

In this paper, we developed a new framework for coherent SAR image formation. This task is challenging due to the problem size and the speckle phenomenon. Current methods also lack uncertainty quantification. Our framework uses a hierarchical Bayesian model with conjugate priors to directly incorporate fully-developed speckle. A parameter-free sparsity-encouraging sampling method is introduced to provide estimates of the image, the speckle, and the noise. The GOTCHA data set examples demonstrates that our method reduces speckle and noise significantly more than other commonly used methods in real world problems. Uncertainty quantification information unprecedented in SAR is also provided in the form of variance images and sample movies, indicating when the pixel values and features shown in an estimate can be trusted. Uncertainty quantification is also provided for the speckle and noise. Such information is of particular importance in SAR, where ground truth images even for synthetically-created examples are unknown.

Future work will focus on further accelerating the sampling method, as well as decreasing storage and memory requirements. This will enable image formation with more pixels, as well as multi-pass and three-dimensional imaging. Finally, comparisons with deep learning based SAR image formation (still in its nascent stages, \cite{mason2017deep}) will be necessary. 

\ifCLASSOPTIONcaptionsoff
  \newpage
\fi



\bibliographystyle{IEEEtran}
\bibliography{refs.bib}
%



%

\begin{IEEEbiographynophoto}{Victor Churchill}
is a Ph.D. student in the Department of Mathematics at Dartmouth College.
\end{IEEEbiographynophoto}
\begin{IEEEbiographynophoto}{Anne Gelb}
is the John G. Kemeny Parents Professor of Mathematics at Dartmouth College.
\end{IEEEbiographynophoto}
\vfill





\end{document}